\documentclass[summary]{ursi}

\usepackage{float}
\usepackage{comment}
\usepackage[table]{xcolor}
\usepackage{cite}      
\usepackage{graphicx}  
\usepackage{psfrag}    
\usepackage{placeins}
\usepackage{lineno}
\usepackage{url}

\usepackage{subcaption}
\usepackage{caption} 
\usepackage{times}
\usepackage{textcomp}
\usepackage{amsmath} 
\usepackage{amsfonts,mathtools}
\usepackage{balance}
\usepackage[normalem]{ulem}
\usepackage{xcolor}

\usepackage{cuted}
\usepackage{amsmath,bm}
\usepackage{hyperref}
\usepackage{cleveref}

\title{Motion Compensation for Multiple-Input-Multiple-Output Inverse Synthetic Aperture Imaging of Automotive Targets}

\author{Devansh Mathur\affref{ref1}, Akanksha Sneh*\affref{ref1}, Debojyoti Sarkar,
  and Shobha Sundar Ram}

\affiliation{%
 {Indraprastha Institute of Information Technology Delhi, New Delhi 110020 India\\

Email: \{devansh21145, akankshas, debojyoti21144, shobha\}@iiitd.ac.in\\
 \aff{ref1}{Authors are joint first authors}}
 
}

\begin{document}

\maketitle
\begin{abstract}
 Inverse synthetic aperture radar (ISAR) images generated from single-channel automotive radar data provide critical information about the shape and size of automotive targets. However, the quality of ISAR images degrades due to road clutter and when translational and higher order rotational motions of the targets are not suitably compensated. One method to enhance the signal-to-clutter-and-noise ratio (SCNR) of the systems is to leverage the advantages of the multiple-input-multiple-output (MIMO) framework available in commercial automotive radars to generate MIMO-ISAR images. While substantial research has been devoted to motion compensation of single-channel ISAR images, the effectiveness of these methods for MIMO-ISAR has not been studied extensively. This paper analyzes the performance of three popular motion compensation techniques - entropy minimization, cross-correlation, and phase gradient autofocus - on MIMO-ISAR. The algorithms are evaluated on the measurement data collected using Texas Instruments millimeter-wave MIMO radar. The results indicate that the cross-correlation MOCOMP performs better than the other two MOCOMP algorithms in the MIMO configuration, with an overall improvement of 36\%.

\end{abstract}

\section{Introduction}
\label{sec:Introduction}
Automotive radars play a vital role in advanced driver assistance systems (ADAS) to enable road safety and avoid road congestion \cite{liu2021data}.
Wideband millimeter-wave (mmW) automotive radars are specifically useful for detecting and identifying different road users based on high-resolution images. Recent advances in inverse synthetic aperture radar (ISAR) imaging of automotive vehicles generated from single channel radar data have utilized the wide-angle perspectives from turning vehicles to capture images of bikes, cars, buses, and trucks \cite{li2015wide}. These ISAR images provide detailed information about each target’s structure, including its shape, size, number of wheels, and trajectory \cite{pandey2022classification}. However, the image quality is affected by radar receiver noise, environmental/road clutter, and errors in the ego vehicle's estimation of its kinematic parameters.

Multiple-input-multiple-output (MIMO) radar systems have been extensively researched as a method for boosting signal-to-clutter-and-noise ratio (SCNR) and enhancing the radar operating metrics \cite{xiong2017fda}. 
In a MIMO radar configuration, the multiple radar antennas at the transmitter and receiver may be widely separated in space (multistatic) or co-located (monostatic). 
In the monostatic configuration, it is possible to precisely phase synchronize the transmitter and receiver antennas and create a virtual array with a large electrical aperture for generating directional beams with fewer elements \cite{xu2015joint}. Several commercial millimeter wave automotive radars support MIMO configurations for different applications.
This work employs a monostatic MIMO automotive radar configuration with time-domain multiplexed (TDM) mode for MIMO, where each transmitting element is activated sequentially through a controlled switching mechanism, 
to generate high-resolution MIMO-ISAR images of automotive targets. 

A significant challenge in achieving good quality high-resolution ISAR images is managing the complex motion dynamics of maneuvering automotive targets. These dynamics include translational motion, such as velocity, acceleration, and jerk, and rotational motion, such as yaw, roll, and pitch. These motion parameters are generally unknown to the radar engineer and distort and blur the radar images. To address these challenges, motion compensation (MOCOMP) has been extensively explored in ISAR literature \cite{itoh1996motion,xing2004migration}. MOCOMP involves estimating the translational and higher-order rotational motion parameters and correcting for their undesired effects on ISAR images. 
Generally, ISAR imaging involves multiple stages of MOCOMP. The first step generally involves \emph{coarse} MOCOMP to mitigate the \emph{range walk} phenomena by phase-correcting for the change in target position across different frames in the raw data \cite{zhang2013translational}. The migration of scatterers across different range cells during the coherent integration time results in Doppler shifts spanning multiple cells. The second stage of MOCAMP involves making these Doppler shifts constant across the cells. This process is commonly referred to as \emph{fine} MOCOMP \cite{gu2004migration}. In some systems, a third stage of MOCAMP is further introduced directly on the image data (rather than raw data) to improve the precision and clarity of the ISAR image. However, all of these works limit their discussion to ISAR images generated with a single transmitter and receiver, and their effectiveness with respect to MIMO-ISAR images has not been extensively studied.

In this work, we qualitatively and quantitatively analyze the performance of three popular  MOCOMP algorithms: entropy minimization, cross-correlation, and phase gradient autofocus, for MIMO-ISAR images. We consider 
real ISAR images generated from measurement data collected with the Texas Instruments AWR1843 millimeter wave MIMO radar system for experimental validation. Our results show that the cross-correlation algorithm is most effective in terms of improving the image quality metric of MIMO-ISAR images. However, it does not perform efficiently for all time instants.

\emph{Notation:} Scalar variables, vectors, and matrices are denoted by small letters in regular font, small letters in bold, and capital letters in bold font, respectively. Symbol $\ast$ represents the convolution operation.

\section{Signal Model and Motion Compensation Framework}
\subsection{Transmit and Received Signal}
\label{sec:Theory}
In this work, we consider a monostatic MIMO configuration consisting of $P$ transmitting antennas and $Q$ receiving antenna elements arranged as a uniform linear array (ULA), 
The transmitting and receiving elements are spaced $d_{tx}$ and $d_{rx}$ distance apart, respectively. 
The transmitting antennas operate sequentially. The receiver includes $Q$ parallel RF hardware chains and is, therefore, capable of digital beamforming. At the transmitter, we generate a frequency-modulated continuous wave (FMCW) signal characterized by a chirp factor $K$ and a pulse repetition interval $T_{PRI}$ as shown 
\par\noindent\small
\begin{align}
 \mathbf{x}(\tau) = \text{rect}\left(\frac{\tau}{T_{PRI}}\right)e^{j\pi K \tau^2}.
 \end{align}
 \normalsize
The signal is then upconverted to a mmW frequency and transmitted sequentially from each $p^{th}$ antenna, as shown  
\par\noindent\small
\begin{align}
\label{eq:TxSig}
\mathbf{x}_{p}(\tau) =\left(\mathbf{x}(\tau)\ast \delta(\tau-(p-1)T_{PRI})\right)
e^{j2\pi f_c \tau}.
\end{align}\normalsize
Here, $\delta(\cdot)$ is the Dirac-delta function and $f_c$ is the millimeter wave carrier frequency (with $\lambda$ wavelength).
The total duration of $P$ transmissions comprises one chirp loop interval ($T_{CLI}$).
Further, $L$ samples of $T_{CLI}$ correspond to the slow time samples, $t$, and form a single coherent pulse interval $T_{CPI}$. 
We assume $B$ scattering centers on an extended target to be present in the radar channel. Each $b^{th}$ scatterer is at a time-varying range of $r_b(t)$ and azimuth $\phi_b(t)$ with respect to the radar. Since we have considered a monostatic radar configuration, the scattering strength, $\sigma_b$(t), is assumed to be identical across all the transmitter-receiver pairs and fluctuates slowly with time. The transmitted signal gets scattered by the targets, and the corresponding reflected echoes are collected and processed at the receiver elements simultaneously. The received signal at each $q^{th}$ element due to $p^{th}$ transmitter, $\mathbf{Y}_{p,q}$, is down-converted and expressed as a radar rectangle of fast time ($\tau$) and slow time ($t$) samples, as shown in
\par\noindent\small
\begin{align}
\nonumber \mathbf{Y}_{p,q}(\tau,t) = \sum_{b=1}^B \sigma_b(t) u_{p,b(t)}u_{q,b(t)}\mathbf{x}_{p}\left(\tau-\frac{2r_b(t)}{c}\right)e^{j2\pi f_{D{_b}}t}.
\end{align}\normalsize
Here, $u_{p,b(t)} = e^{-j\frac{2\pi}{\lambda}d_{tx}(p-1)\sin\phi_b(t)}$ and $u_{q,b(t)} = e^{-j\frac{2\pi}{\lambda}d_{rx}(q-1)\sin\phi_b(t)}$ and $f_{D_b}$ is the Doppler frequency due to the motion of the target. We perform stretch processing by multiplying the received signal with $e^{-jK(\tau-\tau_0)^2}$ to obtain
\par\noindent\small
\begin{align}
\label{eq:stretch_process}
\nonumber \mathbf{\tilde{Y}}_{p,q}(\tau,t) = \sum_{b=1}^B \sigma_b u_{p,b(t)}u_{q,b(t)}e^{-j2\pi f_c \frac{2R_b}{c}} e^{-j \pi K (\delta \tau_b)^2}e^{-j 2\pi \tau_o \delta \tau_b}\\
e^{j2\pi f_{D{_b}}t}e^{-j2\pi K \delta \tau_b \tau}.
\end{align}\normalsize
Here, $\tau_0$ is the delay corresponding to a fixed reference range and
$R_b$ is the initial distance of the $b^{th}$ scatterer with respect to the radar. The time delay to $b^{th}$ point scatterer, $\frac{2r_b(t)}{c}$, is expanded as $(\tau_o+\delta \tau_b)$, where $\delta \tau_b$ is the additional time delay from the reference position. The first three exponential terms in Eq.~\ref{eq:stretch_process} are constant phase terms that can be absorbed into $\sigma_b$. The fourth and fifth exponential terms are linear phase functions of slow time and fast time, respectively.

\subsection{Radar Signal Processing}
The stretch-processed signal is followed by a two-dimensional (2D) Fourier transform, 
to generate the respective range-Doppler ambiguity diagrams. 
This process is repeated for every pair of $p,q$ elements. Then, the $P \times Q$ images are non-coherently integrated to obtain MIMO-ISAR plots for each $T_{CPI}$ throughout the target's motion.
This method enhances the target imaging by leveraging the combined spatial and time domain data from multiple channels, providing detailed visualization of the scatterers. However, the MIMO-ISAR images still contain motion errors due to the translational and rotational components of the target's motion. The range  $r_b(t)$ of $b^{th}$ scatterer comprises the translational and rotational components and can be expanded as 
\par\noindent\small
\begin{align}
\label{eq:motion_comp}
r_b(t) = R(t)+ x_b cos\psi(t)- y_bsin\psi(t). 
\end{align}\normalsize
Here, $R(t)$ corresponds to the range of the target's center of gravity with respect to the monostatic radar. $x_b$ and $y_b$ are the local displacements of the b$^{th}$ scatterer in $x$ and $y$ coordinates respectively from the target's center of gravity. $\psi(t)$ represents the rotational angle that the target is undergoing, which can be further expanded as $\psi(t) = \psi_o + \alpha t +\beta t^2 + \cdots$, where $\psi_o$, $\alpha$ and $\beta$ represent the initial angle, angular velocity and angular acceleration of the target respectively. The ISAR images are generated from the first-order rotational motion after compensating for the translational and higher-order rotational motions. 
\subsection{Motion Compensation Framework}
We perform the coarse and fine MOCOMP on the received signal across $P \times Q$ channels to reduce the image quality degradation. These steps are introduced directly at the raw data stage for each $(p,q)^{th}$ channel data to generate the single input single output (SISO) ISAR image. Then, the images obtained after MOCOMP are non-coherently integrated to obtain MIMO-ISAR. This work ignores the third stage MOCOMP, which is sometimes performed directly on processed images. The coarse MOCOMP is an essential pre-processing step to reduce the dominant phase error due to the translational motion of the target.
Here, a phase correction is introduced to the raw radar data based on the phase change of the highest strength scatterer across the slow-time samples. As a result of the phase correction, the target is fixed in the fast time axis for different slow time intervals. Further, we perform fine MOCOMP to estimate critical motion parameters and eliminate motion effects, thereby improving the clarity of the ISAR images. We consider three popular algorithms for fine, fine MPCOMP: entropy minimization (EM), cross-correlation (CCR), and phase gradient autofocus (PGA). The first method,
EM MOCOMP  \cite{wang2019noise} seeks to minimize entropy by iteratively applying different values for fine motion parameters, velocity, and acceleration in a phase-compensating term to remove the motion effects from the coarse compensated received signal.
Next, the CCR MOCOMP \cite{ozdemir2021inverse} compares successive reflected echoes across multiple channels by calculating the cross-correlation between them, which measures how similar or aligned the signals are across different time intervals.
Further, the PGA MOCOMP method \cite{evers2019generalized} estimates and corrects for the phase difference between the consecutive chirps iteratively, thereby sharpening the ISAR images.
\section{Experimental Setup}
\label{sec:Setup}
\begin{figure}[htbp]
    \centering
    \includegraphics[scale = 0.15]{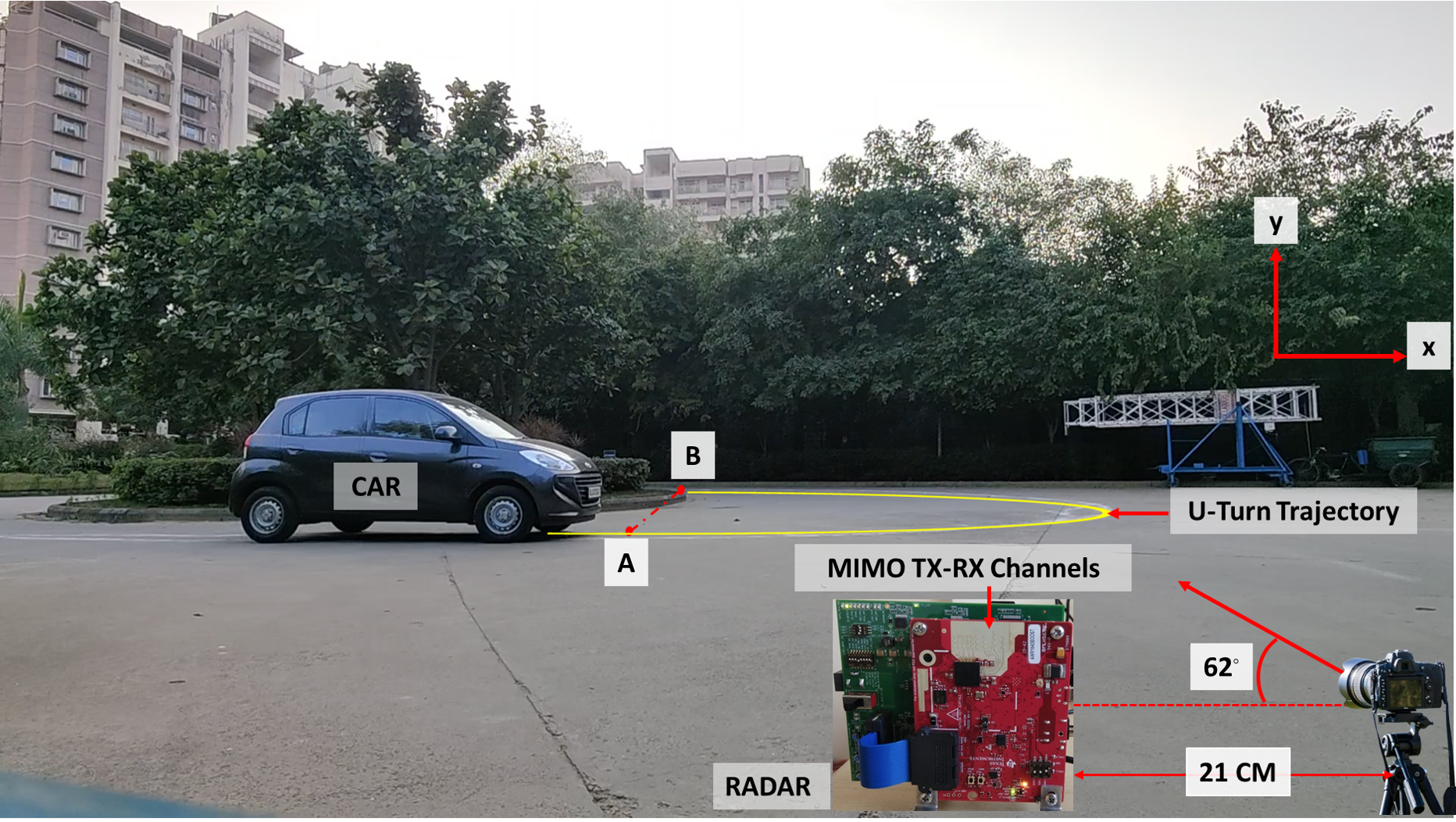}
    \caption{\footnotesize Measurement setup with TI AWR1843 radar in MIMO configuration, the camera (for ground truth information), and a mid-size car as a target.}
    \label{fig:exp_setup}
\end{figure}

    In this work, we describe the experimental setup that involves measurements from TI AWR1843 millimeter wave radar. We consider a Cartesian coordinate system with the ground aligned with the $xy$ plane and the height axis along $z$. The radar is fixed at the origin with a camera fixed at $(0.2,0)$ m to facilitate the recording of ground truth information, as shown in Fig.~\ref{fig:exp_setup}. The radar is oriented towards the positive $y$ axis while the camera is oriented at an angle of 62$^{\circ}$ from the negative $x$ axis. Further, we consider a compact-size car of Hyundai Santro of $3.6 \times 1.6 \times 1.6$ m dimensions for an automotive target. 

The car moves along a U-turn trajectory from an initial position at point A with coordinates $(-14.3,4.9)$ m to point B with coordinates $(-14,30.3)$ m in a duration of 15 s. The rest of the parameters for the measurement setup are listed in Table.\ref{tab:RadarParam_mes}.  
\begin{table}[htbp]
    \centering
    \scriptsize
    \caption{\footnotesize Measurement radar parameters}
    \begin{tabular}{p{2.3cm}|p{1cm}|p{2.5cm}|p{1.2cm}}
    \hline \hline
        \textbf{Configured parameters} & \textbf{Values} & \textbf{Derived parameters} & \textbf{Values}  \\
    \hline \hline
       Carrier frequency  & 77 GHz  & Duty Cycle & 24.6\%  \\
        Radar bandwidth & 2 GHz & Active-Ramp Duty Cycle  & 18.2\% $\mu$ s \\
       Number of Slow Time Samples & 128 & Sampling Frequency  & 9668 ksps \\
       Number of Fast Time Samples & 256 & Maximum unambiguous range & 34.4 m \\
       Coherent processing interval & 0.1 s & Maximum unambiguous velocity & 5 m/s \\
Doppler resolution & 10 Hz & & \\
Minimum cross-range resolution & 0.19 m & & \\
    \hline \hline
    \end{tabular}
    \label{tab:RadarParam_mes}
    \vspace{-2mm}
\end{table}

\section{Results}
\label{sec:Results}
\begin{figure}[htbp]
    \centering
    \includegraphics[scale = 0.22]{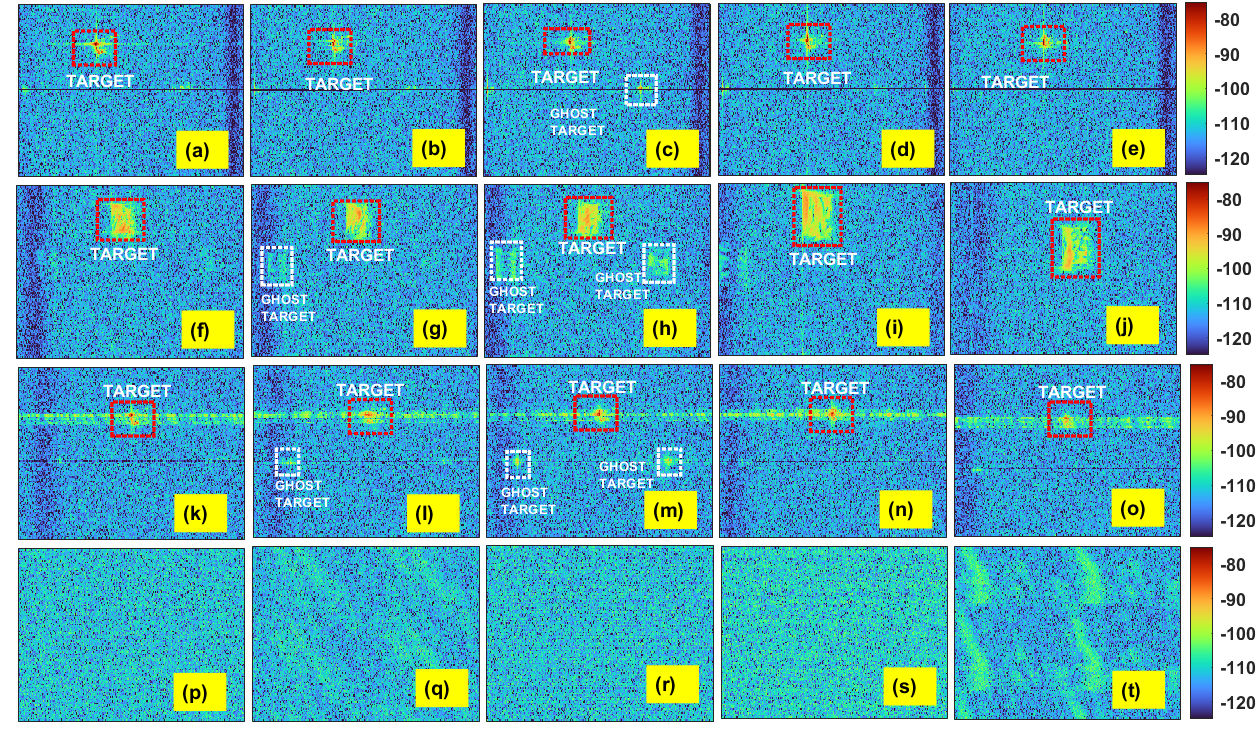}
    \caption{\footnotesize \textcolor{black}{Range-Doppler ambiguity diagrams of the mid-size car taken from radar in SISO configuration where five columns of each row correspond to different time frames: 8.1s, 8.4s, 8.6s, 8.7 and 9.0s. This first row (a-e), second row (f-j), third row (k-o), and fourth row (p-t) includes the plots without MOCOMP, after coarse MOCOMP along with entropy minimization MOCOMP, PGA MOCOMP, and CCR MOCOMP, respectively. The range along the vertical axis spans from 0 to 34.4m with a range resolution of 0.13m, while the Doppler index spans along the horizontal axis spans from 1 to 128.}}
    \label{fig:SISO_meas}
\end{figure}

\begin{figure}[htbp]
    \centering
    \includegraphics[scale = 0.22]{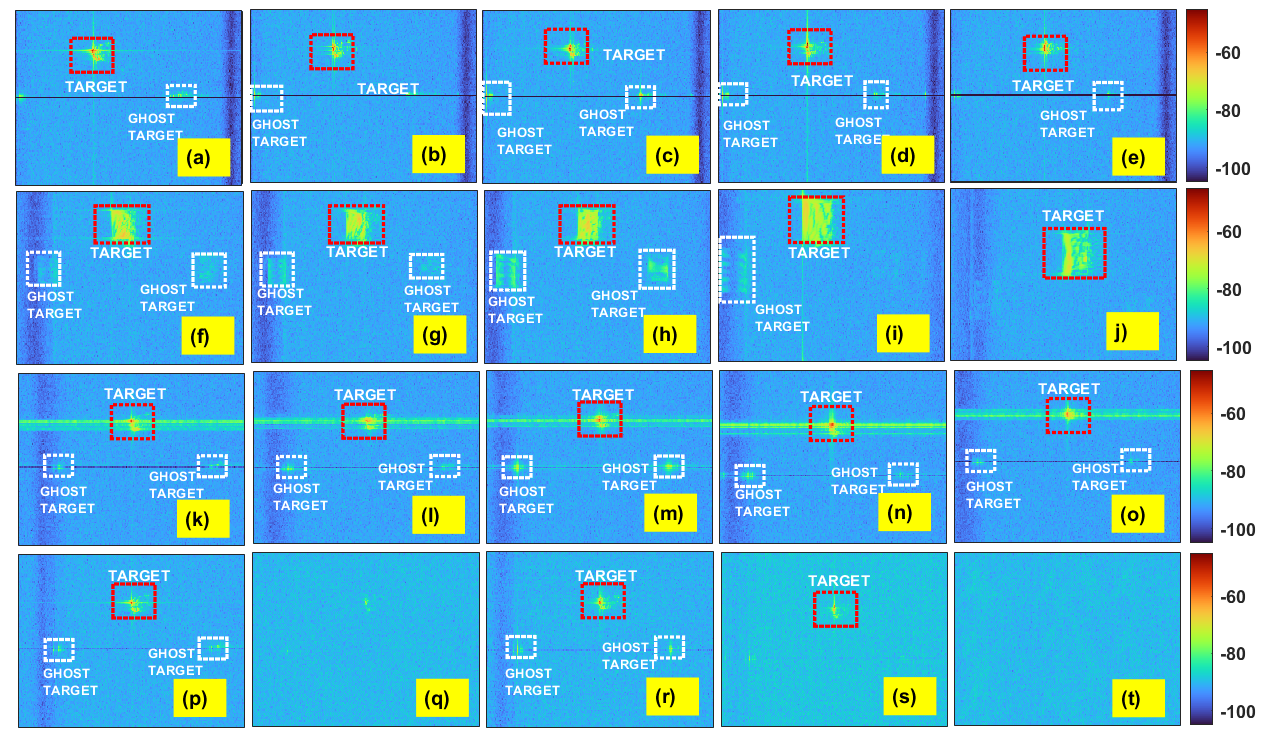}
    \caption{\footnotesize \textcolor{black}{Range-Doppler ambiguity diagrams of the mid-size car taken from radar in MIMO configuration where five columns of each row correspond to different time frames: 8.1s, 8.4s, 8.6s, 8.7 and 9.0s. This first row (a-e), second row (f-j), third row (k-o), and fourth row (p-t) includes the plots without MOCOMP, after coarse MOCOMP along with entropy minimization MOCOMP, PGA MOCOMP, and CCR MOCOMP, respectively. The range along the vertical axis spans from 0 to 34.4m with a range resolution of 0.13m, while the Doppler index spans along the horizontal axis spans from 1 to 128.}}
    \label{fig:MIMO_measured}
\end{figure}

In this section, we discuss the qualitative and quantitative performance of the MOCOMP algorithms. We present the results of the measured data from multiple frames corresponding to different time instants of a car moving along the U-turn trajectory with SISO and MIMO configurations of the radar. In the MIMO setup, 12 MIMO ISAR images are generated for each frame, corresponding to all transmitter-receiver channel combinations. These images are subsequently non-coherently integrated to produce a MIMO-ISAR consolidated frame for each time instant. Firstly, we discuss the results for SISO configuration in Fig.~\ref{fig:SISO_meas}, where each column represents a frame of time instants: 8.1s, 8.4s, 8.6s, 8.7s, and 9.0s. These time instants are chosen as these frames capture the car's turning motion along the U-trajectory and are most useful for generating ISAR images. Further, we obtain some blank frames at initial time instants for both SISO and MIMO configurations where the target is not in the radar's field of view. These frames are used for calculating the noise floor.  
All the results for the SISO case show that poor SCNR limits the image quality. 
The first row shows ISAR images without MOCOMP. Here, we observe the displacement of the target along the range axis for different frames due to the translational motion of the car. Further, we observe prominent fluctuations in the noise floor. We present the results for ISAR images after coarse MOCOMP in Fig.~\ref{fig:SISO_meas} (f-t) for multiple frames and observe that the target gets fixed along the range and Doppler axes. In Fig.~\ref{fig:SISO_meas} (f-j), ISAR results after fine MOCOMP based on entropy minimization are presented, where we observe significant smearing of the radar signatures. However, the noise floor fluctuations have slightly improved along with the clutter suppression. The results from the measurement data are, therefore, significantly different from the simulation scenario, indicating the limitations of the algorithm's performance in real-world scenarios. 
The results for PGA MOCOMP are presented in Fig.~\ref{fig:SISO_meas} (k-o), where we observe a significant improvement in the radar signatures as compared to the entropy minimization but the range sidelobes are prominently visible in all the frames. Lastly, we present the results for CCR MOCOMP in Fig.~\ref{fig:SISO_meas} (p-t). This algorithm is implemented iteratively with slight improvement in each iteration. In this work, we consider 97 iterations to obtain efficient MOCOMP results. However, this algorithm performs very well for some frames (u, w, x) but does not for the v and y frames of Fig.~\ref{fig:SISO_meas}. Next, we present the results for the MIMO configuration in Fig.~\ref{fig:MIMO_measured}. 
Here, we observe that the SCNR of the MIMO-ISAR images is greater than the SISO-ISAR images for all the frames capturing the car's motion. The first row includes the results generated without MOCOMP. As a result, we observe the displacements due to the translational motion. The rest of the rows represent the viewgraphs with coarse MOCOMP, where we observe that the target gets fixed along the range axis and Doppler axis. The second row shows the results with fine MOCOMP based on entropy minimization. Here, the targets are smeared/blurred for every time frame. The third row of Fig.~\ref{fig:MIMO_measured} shows the results generated from PGA-based fine MOCOMP. Here, we observe the sidelobes across the Doppler axis become prominent. Lastly, we present results with CCR MOCOMP in the fourth row, where we observe significant improvement in MIMO ISAR images of frames (p),(r), and (s). However, it performs poorly for frames (q) and (t). The measurement results presented in Fig.~\ref{fig:SISO_meas} and Fig.~\ref{fig:MIMO_measured} also show the presence of ghost targets attributed to multipath effects. We also compare the MOCOMP algorithms through a quantitative metric - the coefficient of variation - of the noise floor for both sets of images in Table.\ref{tab:ResultsMetrics}. This metric is the ratio between the mean and variance of the noise floor calculated from the blank frames obtained from initial time instants. The metric accounts for both SCNR improvement in the MIMO configuration versus the SISO and the focusing capabilities of the fine MOCOMP algorithms. The table shows that this value is lower for the MIMO-ISAR for all MOCOMP cases than for the SISO. This results in an overall improvement in the contrast between the signal returns from the target and the noise and clutter components in the background. The results of fine MOCOMP with CCR are superior to PGA and entropy minimization, demonstrating an overall improvement of 36\% for MIMO-ISAR.
\begin{table}
\centering
\scriptsize
\caption{\footnotesize Coefficient of variation in MIMO and SISO for different MOCOMP algorithms applied on the measured data}
\label{tab:ResultsMetrics}
\begin{tabular}{p{2.5cm}|p{1cm}|p{1cm}|p{1cm}|p{1cm}}
\hline\hline
\textbf{Case} & \textbf{SISO} & \textbf{MIMO}&\textbf{\% improvement in SISO}&\textbf{\% improvement in MIMO}\\
\hline\hline
No MOCOMP & 0.56 & 0.24&-&-\\
Entropy minimization& 0.56 & 0.24&0.1&0.1\\
Phase gradient autofocus & 0.54 & 0.19&3.57&17.19\\
Cross-correlation & 0.52 & 0.15&6.76&36.26\\
\hline\hline
\end{tabular}
\end{table}

\section{Conclusion}
\label{sec:Conclusion}
This work presents MOCOMP MIMO-ISAR radar imaging results for measured experiments at millimeter-wave frequencies. The MIMO-ISAR images demonstrated enhanced signal strength compared to their SISO-ISAR counterparts. This is especially useful in automotive scenarios where the images are characterized by significant road clutter. Among the motion compensation techniques, CCR MOCOMP outperformed both entropy minimization and PGA, achieving a 36\% overall improvement in image quality for MIMO-ISAR. However, this method was not consistently effective for all frames. These results underscore the requirement for new algorithms for MOCOMP in MIMO scenarios for real-world conditions and diverse target scenarios.

\bibliographystyle{ieeetran}
\bibliography{Bibliography}

@article{pandey2022classification,
  title={Classification of automotive targets using inverse synthetic aperture radar images},
  author={Pandey, Neeraj and Ram, Shobha Sundar},
  journal={IEEE Transactions on Intelligent Vehicles},
  volume={7},
  number={3},
  pages={675--689},
  year={2022},
  publisher={IEEE}
}

@article{xiong2017fda,
  title={FDA-MIMO radar range--angle estimation: CRLB, MSE, and resolution analysis},
  author={Xiong, Jie and Wang, Wen-Qin and Gao, Kuandong},
  journal={IEEE Transactions on Aerospace and Electronic Systems},
  volume={54},
  number={1},
  pages={284--294},
  year={2017},
  publisher={IEEE}
}

@article{xu2015joint,
  title={Joint range and angle estimation using MIMO radar with frequency diverse array},
  author={Xu, Jingwei and Liao, Guisheng and Zhu, Shengqi and Huang, Lei and So, Hing Cheung},
  journal={IEEE Transactions on Signal Processing},
  volume={63},
  number={13},
  pages={3396--3410},
  year={2015},
  publisher={IEEE}
}

@book{ozdemir2021inverse,
  title={Inverse synthetic aperture radar imaging with MATLAB algorithms},
  author={Ozdemir, Caner},
  year={2021},
  publisher={John Wiley \& Sons}
}

@inproceedings{li2015wide,
  title={Wide-angle ISAR imaging of vehicles},
  author={Li, Chenchen J and Ling, Hao},
  booktitle={2015 9th European Conference on Antennas and Propagation (EuCAP)},
  pages={1--2},
  year={2015},
  organization={IEEE}
}

@article{liu2021data,
  title={A data fusion model for millimeter-wave radar and vision sensor in advanced driving assistance system},
  author={Liu, Yang and Liu, Yan},
  journal={International Journal of Automotive Technology},
  volume={22},
  number={6},
  pages={1695--1709},
  year={2021},
  publisher={Springer}
}

@article{xing2004migration,
  title={Migration through resolution cell compensation in ISAR imaging},
  author={Xing, Mengdao and Wu, Renbiao and Lan, Jinqiao and Bao, Zheng},
  journal={IEEE Geoscience and Remote Sensing Letters},
  volume={1},
  number={2},
  pages={141--144},
  year={2004},
  publisher={IEEE}
}

@article{itoh1996motion,
  title={Motion compensation for ISAR via centroid tracking},
  author={Itoh, Toshiharu and Sueda, Hachiro and Watanabe, Ysuo},
  journal={IEEE Transactions on Aerospace and Electronic Systems},
  volume={32},
  number={3},
  pages={1191--1197},
  year={1996},
  publisher={IEEE}
}

@article{zhang2013translational,
  title={Translational motion compensation for ISAR imaging under low SNR by minimum entropy},
  author={Zhang, Lei and Sheng, Jia-lian and Duan, Jia and Xing, Meng-dao and Qiao, Zhi-jun and Bao, Zheng},
  journal={EURASIP Journal on Advances in Signal Processing},
  volume={2013},
  pages={1--19},
  year={2013},
  publisher={Springer}
}

@article{evers2019generalized,
  title={A generalized phase gradient autofocus algorithm},
  author={Evers, Aaron and Jackson, Julie Ann},
  journal={IEEE Transactions on Computational Imaging},
  volume={5},
  number={4},
  pages={606--619},
  year={2019},
  publisher={IEEE}
}

@article{gu2004migration,
  title={Migration based SAR imaging for ground penetrating radar systems},
  author={Gu, Kunlong and Wang, Gang and Li, Jian},
  journal={IEE Proceedings-Radar, Sonar and Navigation},
  volume={151},
  number={5},
  pages={317--325},
  year={2004},
  publisher={IET}
}

@article{wang2019noise,
  title={Noise-robust motion compensation for aerial maneuvering target ISAR imaging by parametric minimum entropy optimization},
  author={Wang, Jiadong and Zhang, Lei and Du, Lan and Yang, Dongwen and Chen, Bo},
  journal={IEEE Transactions on Geoscience and Remote Sensing},
  volume={57},
  number={7},
  pages={4202--4217},
  year={2019},
  publisher={IEEE}
}
\end{document}